\documentclass[%
twocolumn,
amsmath,amssymb,
aps,
prl,
]{revtex4-2}
\usepackage{graphicx}
\usepackage{dcolumn}
\usepackage{bm}
\usepackage{hyperref}
\usepackage{amsmath}
\usepackage{amssymb}
\newcommand{\ds}{\displaystyle}
\newcommand{\beq}{\begin{equation}}
	\newcommand{\eeq}{\end{equation}}
\newcommand{\beqq}{\begin{equation*}}
	\newcommand{\eeqq}{\end{equation*}}
\newcommand{\beqa}{\begin{eqnarray}}
	\newcommand{\eeqa}{\end{eqnarray}}

\usepackage{xcolor}

\begin{document}
%%%%%%%%%%%%%%%%%%%%%%%%%%%%%%%%%%%%%%%%%%
\title{{Escape from an attractor generated by recurrent exit}}
\author{Lou Zonca$^{1,2}$ and David Holcman$^{1}$}
\affiliation{$^{1}$Group of Data modeling, Computational Biology and Applied Mathematics,
Ecole Normale Sup\'erieure-PSL, 75005 Paris, France.\\ $^{2}$ Sorbonne University, Pierre et Marie Curie Campus, 75005 Paris, France.}
%%%%%%%%%%%%%%%%%%%%%%%%%%%%%%%%%%%%%%%%%%
\begin{abstract}
Kramer's theory of activation over a potential barrier consists in computing the mean exit time
from the boundary of a basin of attraction of a randomly perturbed dynamical system. Here we report that for some systems, crossing the boundary is not enough, because stochastic trajectories return inside the basin with a high probability a certain number of times before escaping far away. This situation is due to a shallow potential. We compute the mean and distribution of escape times and show how this result explains the large distribution of interburst durations in neuronal networks.
\end{abstract}
\maketitle
\begin{figure}
	\includegraphics[scale=0.43]{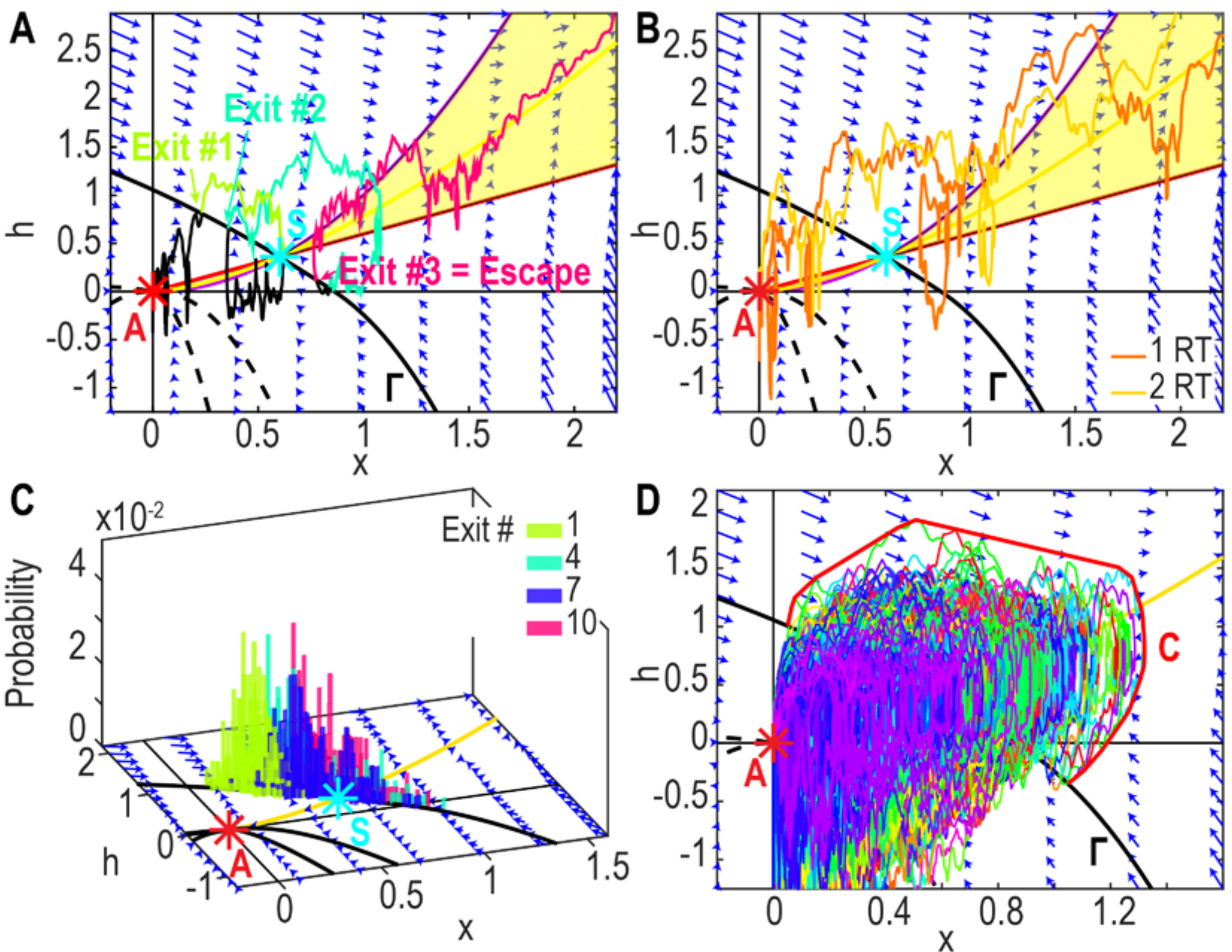}
	\caption{\textbf{Recurrent escape patterns} \textbf{A.} Escaping trajectories reach
the separatrix $\Gamma$ for the first time (step 1, black) and re-cross it several times going back and forth inside and outside of the basin of attraction (step 2, green, cyan, blue) before eventually escaping far away (pink). \textbf{B.} Stochastic trajectories with one (yellow) and two (orange) round-trip (RT) before escape. \textbf{C.}	Distributions of exit points on $\Gamma$ (500 runs) for successive RT. \textbf{D.} Outer boundary layer $C$ computed as the convex hull of all trajectories reentering the basin of attraction (red).}\label{fig1}
\end{figure}
\begin{figure}
	\includegraphics[scale=0.43]{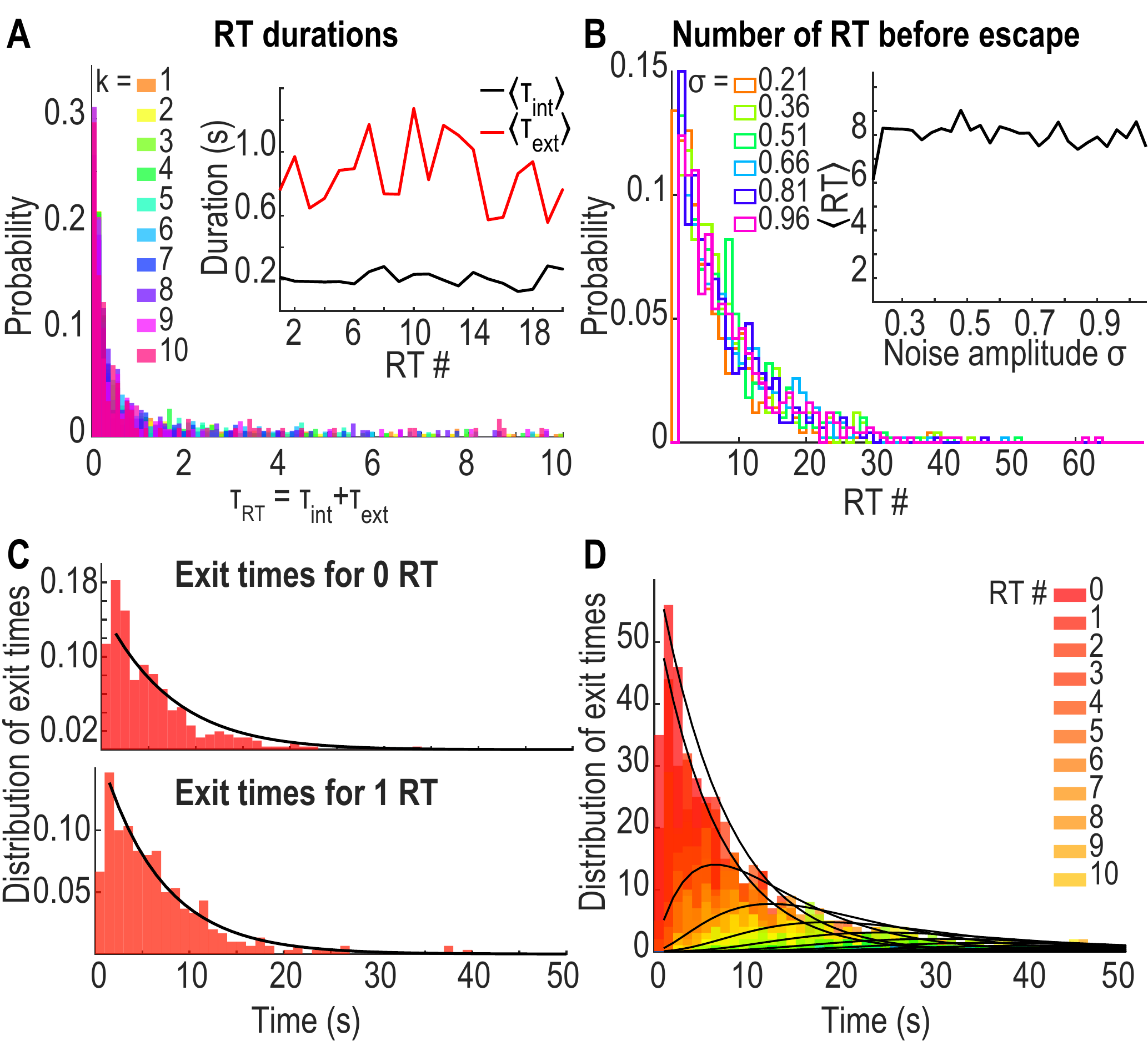}
	\caption{\textbf{Distribution of RT and escape times. }\textbf{A.} Distributions of a
RT duration $\tau_{RT,k}= \tau_{ext,k}+\tau_{int,k}$ for $k \in [1,10]$. Inset: mean time spent outside (resp. inside) the basin of attraction $\langle\tau_{ext,k}\rangle$ (resp. $\langle\tau_{int,k}\rangle$) vs the RT number $k$. \textbf{B. }Distributions of the RT number around the separatrix before a trajectory eventually escapes for various values of $\sigma$ (with $\gamma=0.6$ and $\alpha=1$). 500 runs for each value of $\sigma$. Inset: mean RT number with respect to the noise amplitude $\sigma$. \textbf{C.} Distributions $f_0$ (upper), resp. $f_1$ (lower), of escape times for trajectories with zero and one RT.  The fit uses eq. \eqref{fitFi}. \textbf{D.} Distribution of exit times with the contribution of each RT number compared to the analytical distribution (eq. \ref{pdfExitTimes}).}\label{fig2}
\end{figure}
\begin{figure}
	\includegraphics[scale=0.43]{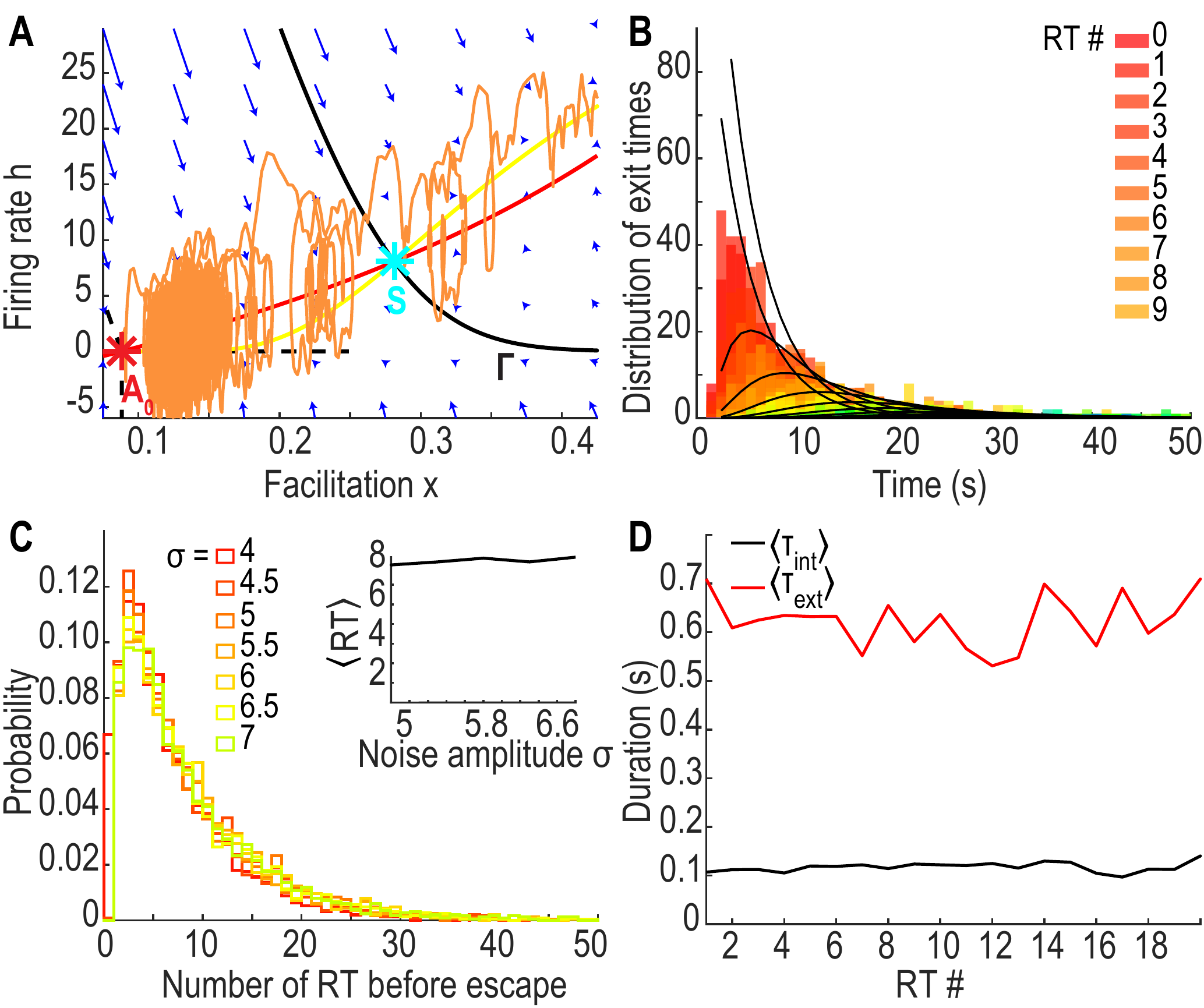}
	\caption{\textbf{Application to the dynamical system \eqref{2Dsyst}. }\textbf{A.} 2D
phase-space restricted to $\{x \leq 0.5 \mbox{ and } h \leq 30\}$. The basin of attraction of $A_0$ (yellow star) is delimited by the stable manifold of $S$ (solid black curve $\Gamma$) with an exiting trajectory doing 1 RT (orange) around the separatrix before escape . \textbf{B.} Distribution of exit times with the contribution of the trajectories per RT number before escape with the analytical fit (equations \ref{pdfExitTimes}, \ref{f0bursts} and \ref{f1bursts}). \textbf{C.} Distribution of the RT number for $\sigma \in [4,7]$ and mean RT number with respect to noise (inset). \textbf{D.} Values of $\langle\tau_{ext}\rangle$ (red) and $\langle\tau_{int}\rangle$ (black) with respect to the RT number.} \label{application}
\end{figure}

In Kramers'theory \cite{Kramers1940,schuss1980,Schuss2010,gardiner1985handbook}, the escape time over a potential barrier consists in computing the mean first passage time (MFPT) of a dynamical system perturbed by a small noise to the boundary of a basin of attraction. The MFPT measures the stability and provides great insight of the backward binding rate in chemistry \cite{dykman1994large,nitzan2006}, loss-of-lock for phase controllers in communication theory \cite{Schuss2011}, escape of receptors from the post-synaptic density at neuronal synapses and is also used to evaluate future derivatives in the financial market \cite{papanicolaou2000}. The full distribution of exit times can be used to characterize both short and intermediate time asymptotics relevant in polymer physics \cite{hawk2013computation}, accelerating chemical reaction simulations \cite{dellago1998efficient}, or better characterizing the search for a small target in a complex environment \cite{godec2016universal,grebenkov2019fullDistrib}.\\
In the limit of small noise, a trajectory escapes a basin of attraction with probability one \cite{Matkowsky1977exit}, but the escape time is exponentially long depending on the topology of the noiseless dynamics \cite{Freidlin1998,smelyanskiy1999time} and its behavior at the boundary. In addition, the distribution of exit points peaks at a distance $O(\sqrt{\sigma})$ from a saddle-point, where $\sigma$ is the noise amplitude \cite{schuss1980,BobrovskySchuss1982,Schuss2011}. Interestingly, when a focus attractor is located near the boundary of the basin of attraction, the escape time deviates from an exponential distribution because trajectories oscillate inside the attractor before escape \cite{verechtchaguina2006_1,verechtchaguina2006_2,verechtchaguina2007,tuckwell2009inhibition,daoduc2016}. \\
In these previous examples, the escape ends at the first time a trajectory crosses the separatrix
that delimits the basin of attraction. Recurrent returns inside a basin of attraction can be quantified by the Green's function of the inner domain used in the additive properties of the MFPT \cite{matkowsky1984}. In their specific case, where the escape time consists in the first crossing of the boundary of the basin of attraction and a second separatrix, their results show a factor two between the escape time and the exit from the basin of attraction. In dimension one, a recurrent return can be quantified using a relaxation time computed from the survival probability when it does not converge to zero in the long times regime \cite{agudov1999}. We show here that for some shallow two-dimensional dynamical systems, trajectories can first exit the basin of attraction, then make excursions outside before coming back inside the domain, a behavior that occurs several times before eventually escaping far away. This situation is peculiar and specific to dimensions greater than two and these recurrent entries need to be taken into account in computing the final escape time.\\
This letter reports such phenomenon. We present formulas for the mean and distribution of escape times and we show that these recurrent re-entries inside the basin of attraction can increase the escape time by a factor between two and three. Finally, we apply these results
to explain the origin of long interburst durations found in neuronal network models \cite{CoombesBressloff}.\\
{\bf Recurrent escape patterns.}
%%%%%%%%%%%%%%%%%%%%%%%%%%%%%%%%%%%%%%%%%%%%%%%%%%%%%%%
We start with a generic two-dimensional system
\beq\label{Drift}
\begin{array}{r c l}
	\arraycolsep=1.4pt\def\arraystretch{2.5}
	\dot{h}&=&-\alpha h + x^2 +\sigma \dot{\omega}\\
	\dot{x} &=& \left\{\begin{array}{l c l}
		\arraycolsep=1.4pt\def\arraystretch{2.5}
		h - \gamma x &\text{for}& h\geq 0\\
		- \gamma x &\text{for}& h\leq 0,\\
	\end{array} \right.
\end{array}
\eeq
where $\alpha \in ]0,1]$, $\gamma \in ]0, \alpha[$, $\dot{\omega}$ is a Gaussian
white noise and $\sigma$ its amplitude. The determinist part of this system has two critical points: one attractor $A = (0,0)$ (fig. \ref{fig1}A red star) and one saddle-point $S = (\gamma^2 \alpha, \gamma \alpha)$ (fig. \ref{fig1}A cyan star) and the separatrix $\Gamma$ delimits the basin of attraction of $A$ (fig. \ref{fig1}A solid black).\\
%%%%%%%%%%%%%%%%%%%%%%%%%%%%%%%%%
%%%%%%%%%%%%%%%%%%%%%%%%%%%%%%%%%%%	
The escape of the basin of attraction occurs in two steps. 1) A trajectory starting at $A$ reaches $\Gamma$ for the first time (fig. \ref{fig1}A black trajectory between $A$ and the first exit point Exit $\#1$, light green). 2) The trajectory exits and crosses $\Gamma$ several times, that we count by using a round-trip (RT) number (fig. \ref{fig1}A, light green and cyan loops) before eventually escaping far away (fig. \ref{fig1}A, pink). To characterize the final escape times and the distribution of crossing points on $\Gamma$, we ran stochastic simulations of system \eqref{Drift} (500 runs) fig. \ref{fig1}B (trajectories exhibit one (yellow) and two (orange) RT before escape). To further characterize the recurrent crossing points, we plotted their distributions (fig. \ref{fig1}C) and found that they were peaked near the saddle-point. These recurrent excursions are not due to a focus, since the saddle point $S$ has only real eigenvalues $\lambda_\pm = -\cfrac{1}{2}\left(-(\alpha+\gamma)\pm \sqrt{(\alpha+\gamma)^2+4\alpha \gamma}\right)$, $\lambda_+ \approx 0.314, \lambda_-\approx -1.914$ (with $\alpha=1$ and $\gamma=0.6$). A possible explanation for this phenomenon is the very shallow field tangent to the separatrix: only near the unstable manifold (fig. \ref{fig1}A yellow curve) trajectories can depart to infinity when there are located inside the ensemble of points where the two drift components are positive $\dot{x}>0$ and $\dot{h}>0$, thus  $C_\infty = \left\{ h>\gamma x \mbox{ and } h<\cfrac{x^2}{\alpha}\right\}$ (fig. \ref{fig1}A, B yellow area, situated between the x-nullcline (red) and the h-nullcline (purple)). Before reaching $C_{\infty}$, the noise pushes the trajectories back and forth into the basin of attraction.\\
To conclude this part we shall summarize the escape dynamics:
\begin{enumerate}
\item The distribution of exit points peaks at a distance $O(\sqrt{\sigma})$ from the saddle-point (generically satisfied \cite{BobrovskySchuss1982}).\\
\item The shallow field near the separatrix allows the trajectories to reenter with high probability.\\
\item The peaks of the successive exit points distributions drift towards the saddle-point $S$ (fig. \ref{fig1}C).\\
\item When the trajectories enter the escape cone $C_{\infty}$ (yellow surface in fig. \ref{fig1}A-B) where the field increases, they eventually escape to infinity.\\
\end{enumerate}
Finally, this escape pattern could not occur in dimension one since conditions 1 and 3 cannot be satisfied.\\
%%%%%%%%%%%%%%%%%%%%%%%%%%%%%%%%%%%%%%%%%%%%%%%%%%%%%%%
{\bf Characterizing the escape time.}
%%%%%%%%%%%%%%%%%%%%%%%%%%%%%%%%%%%%%%%%%%%%%%%%%%%%%%%
We compute here the total escape time. For that goal, we decomposed it into the time to reach
the separatrix $\Gamma$ for the first time plus the time spent to go back and forth around $\Gamma$ before the final escape. Using Baye's law and conditioning on the RT numbers, the mean escape time can be written as
\beq \label{realEscapeTime}
\langle \tau_{esc} \rangle = \sum_{k=0}^{\infty} \langle \tau | k \rangle P_{RT}(k),
\eeq
where $\langle \tau | k \rangle$ (resp. $P_{RT}(k)$) is the mean time (resp. probability) to
return $k$ times inside the basin of attraction. To estimate the escape probability $\tilde{p}$ for a trajectory that had crossed $\Gamma$ to escape to infinity, we ran $N=500$ trajectories starting from $A$ and lasting $T=300s$. We first counted the proportion of trajectories reentering the basin of attraction at least once and obtained 88\%. We then reiterated this process and counted the proportion of trajectories reentering the basin of attraction one more time after each RT. We found that this proportion was stable equal to 88\%, leading to $\tilde{p}=0.12$. We applied this process for values of the noise amplitude $\sigma \in [0.21,1.05]$ and found that $\tilde{p}$ did not depend on $\sigma$. After $T=300s$ all trajectories had escaped to infinity (for all the values of $\sigma$), thus choosing a higher value for $T$ would not change the value of $\tilde{p}$. This escape phenomenon could be interpreted as follows: a trajectory has escaped when it reaches a distance far away from the separatrix and to better characterize such a distance outside the basin of attraction, we generated empirical trajectories that will return (have not yet escaped) and estimated their convex hull $C$ (fig. \ref{fig1}D red, 500 runs). Formally, this is equivalent to looking at trajectories starting at $A$ conditioned to a return to the basin of attraction, thus defining a sort of Brownian bridge. This procedure leads to a bounded domain: any point inside $C$ has a high probability of reentering the basin of attraction while points further away will escape to infinity.\\
Due to the strong Markovian properties, each RT can be considered independent of the previous ones,
thus the probability to escape after exactly $k-$RT is given by
\beq
P_{RT}(k) = \tilde{p}(1-\tilde{p})^{k-1},
\eeq
and thus the mean escape time is
\beq\label{tEscSummed}
\begin{array}{r c  l}
\langle \tau_{esc} \rangle &=& \langle\tau_0\rangle + (\langle\tau_{ext}\rangle+\langle\tau_{int}\rangle)\tilde{p}\sum_{k=1}^\infty k (1-\tilde{p})^{k-1}\\
&=& \langle\tau_0\rangle + \cfrac{\langle\tau_{ext}\rangle+\langle\tau_{int}\rangle}{\tilde{p}},
\end{array}
\eeq
where  $\langle\tau_0\rangle$ is the mean time to reach the separatrix for the first
 time and $\langle\tau_{ext}\rangle$ (resp. $\langle\tau_{int}\rangle$) is the time spent on the outside (resp. inside) the basin of attraction of $A$ for each RT (fig. \ref{fig2}A). When the escape probability $\tilde{p}$ tends to zero, the escape time tends to infinity, corresponding to trajectories that would be trapped in $C$. In our case, the mean escape time is  $\langle\tau_{esc}\rangle \approx \langle\tau_0\rangle + 8.33(\langle\tau_{ext}\rangle+\langle\tau_{int}\rangle)$. With the present parameters $\langle\tau_{0}\rangle \approx 5.1s$ and $\langle\tau_{ext}\rangle+\langle\tau_{int}\rangle \approx 1s$ showing that the escape time is increased by a factor $2.6$. Interestingly, the noise amplitude does not influence the number of RT before escape (fig. \ref{fig2}B). For the parameter value $\gamma=0.6$, we found that a trajectory perform 8 RT on average (fig. \ref{fig2}B, inset). These results indicate that the noise amplitude does not directly influence the probability to escape to infinity.
%%%%%%%%%%%%%%%%%%%%%%%%%%%%%%%%%%%
%%%%%%%%%%%%%%%%%%%%%%%%%%%%%%%%%%%
We now determine the distribution of escape times
\beq
P (\tau_{esc}<t)=\sum_{k=0}^{\infty} P (\tau^k<t|k) P_{RT}(k),
\eeq
where $P(\tau^k<t|k)$ is the conditional probability distribution to escape after $k$ RT.
Because RT are i.i.d, this probability is the $k$-th convolution of the distribution of times of a single RT $f_1(t)$ with the distribution of escape times without RT $f_0(t)$
\beq
P(\tau^k<t|k)=f_0(t)*f_1(t)^{*k},
\eeq
where $f(t)^{*k}=f(t)*f(t)*...*f(t)$, $k$ times. Thus the pdf of exit times is given by
\beq\label{pdfExitTimes}
f(t)=\sum_{k=0}^{\infty} f_0(t)*f_1(t)^{*k} \tilde{p}(1-\tilde{p})^{k-1}.
\eeq
To compare this formula to the results of our numerical simulations, we approximate the
distributions $f_0$ and $f_1$ by
\beq\label{fitFi}
f_i(t)=c_i \left(1+erf\left(\cfrac{t-a_i}{b_i}\right)\right) e^{\ds -\lambda_i t}, \text{ for } i=0,1,
\eeq
where $\ds erf(x)=\cfrac{2}{\sqrt{\pi}}\int_{0}^{x}e^{-u^2}du$ is the error function. We
fitted the distributions obtained from the numerical simulations of trajectories that escaped without doing any RT ($f_0$ fig. \ref{fig2}C, upper) and after one single RT ($f_1$ fig. \ref{fig2}C, lower) with the condition that  $\lambda_1\geq\lambda_0$. We obtained $c_0 = 1.09$, $c_1 = 1.63$, $\lambda_0 = 0.06$ $\lambda_1 = 0.13$, $a_0 = -38.28$, $a_1 = -138.64$, $b_0 = -36.72$, $b_1 = -121.69$. We then computed each term of the sum \eqref{pdfExitTimes} and we could compare it to the corresponding parts of the distribution of escape times obtained from stochastic simulations (fig. \ref{fig2}D).\\
%%%%%%%%%%%%%%%%%%%%%%%%%%%%%%%%%%%%%%%%%%%%%%%%%%%%%%%
{\bf Interburst durations in a firing excitatory neuronal network}
%%%%%%%%%%%%%%%%%%%%%%%%%%%%%%%%%%%%%%%%%%%%%%%%%%%%%%%
Burst and interburst are fundamental network events occurring during dominant imbalance
dominated by excitatory neuronal activity. Network burst generation could rely on specific spiking frequencies in connected neurons \cite{ermentrout2001} despite a high variability in interspike intervals \cite{gutkin1998}. Neuronal population bursts separated by long interbursts have been modeled using a two-state synaptic depression \cite{Guerrier2015}, or by using the refractory period induced by afterhyperpolarization (AHP), a mechanism leading to a long voltage hyperpolarisation transient and generated by various potassium channels \cite{AHPmodel}. Here we show that the recurrent escape mechanism described above can be used as one explanation of the origin of long interburst intervals without the need of any other mechanism. However, we note that this mechanism does not have to be exclusive and that long interburst intervals could also be explained in some cases by a combination of mechanisms such as the recurrent escape pattern presented here and AHP. Indeed, we start from the depression-facilitation short-term synaptic plasticity mean-field model of network neuronal bursting \cite{Tsodyks1997,DaoDuc2015,Holcman_Tsodyks2006}, which consists of three equations \eqref{sys} for the mean voltage $h$, the depression $y$, and the facilitation $x$. The depression mechanism describes the depletion of the vesicular pool necessary for neurotransmission following successive action potentials, while the facilitation mechanism corresponds to a transient increase of the release probability mediated by a local calcium accumulation at synapses.
%%%%%%%%%%%%%%%%%%%%%%%%%%%%%%%%%%%%%%%%%%%%%%%%%%%%
\beqa \label{sys}
\tau \dot{h} &=& - h + Jxy h^+ +\sqrt{\tau}\sigma \dot{\omega}\nonumber\\
\dot{x} &=& \dfrac{X-x}{t_f} + K(1-x) h^+ \\
\dot{y} &=& \dfrac{1-y}{t_r} - L xy h^+ , \nonumber
\eeqa
%%%%%%%%%%%%%%%%%%%%%%%%%%%%%%%%%%%%%%%%%%%%%%%%%%%%%
where $h^+ = max(h,0)$ is a linear threshold function of the synaptic current that gives the average population firing rate \cite{Tsodyks1997,Holcman_Tsodyks2006,Barak2007}. The mean number of connections (synapses) per neuron is accounted for by the parameter $J$ and the term $Jxy$ represents the combined effect of the short-term synaptic plasticity (facilitation and depression mechanisms) on the network activity. The parameters $K$ and $L$ describe how the firing rate is transformed into molecular events that are changing the duration (depression) and probability (facilitation) of vesicular release. The time scales $t_f$ and $t_r$ define the recovery of an averaged synapse from the network activity. Finally, $\dot \omega$ is an additive Gaussian noise and $\sigma$ its amplitude, this additive noise term represents the fluctuations of the mean voltage generated by the average of independent vesicular release events and/or closings and openings of voltage gated channels.\\
This system has 3 critical points, one attractor and two saddles. Interestingly, near the attractor
 $A=(0,X,1)$, the dynamic is anisotropic ($|\lambda_1| = 12.6 \gg |\lambda_2| = 1.11 \gg |\lambda_3| = 0.34$, with the parameters from Table \ref{tableParam}) and thus we project the system on the two-dimensional plan $y=Cte$
\beq
\dot{y}=0=\dfrac{1-y}{\tau_r} - L xy h^+ = 0 \iff y = \cfrac{1}{1+\tau_r Lxh^+}
\eeq
leading to the simplified system
%%%%%%%%%%%%%%%%%%%%%%%%%%%%%%%%%%%%%%%%%%%%%%%%%%%%%
\beq\label{2Dsyst}
\begin{array}{r c l}
	\arraycolsep=1.4pt\def\arraystretch{2.5}
	\dot{h}&=&\cfrac{h\left(Jx-1-\tau_rLxh^+ \right)}{\tau(1+\tau_rLxh^+)}+\sqrt{\tau}\sigma \dot{\omega}\\
	\dot{x} &=& \cfrac{X-x}{\tau_f} + K(1-x) h^+
\end{array}
\eeq
%%%%%%%%%%%%%%%%%%%%%%%%%%%%%%%%%%%%%%%%%%%%%%%%%%%%%
The deterministic component of this system has 3 critical points, two attractors and one saddle-point .
\paragraph{Attractor $A_0$} A first equilibrium point is given by $h=0$ and $x=X$.
The Jacobian at this point is
\beq\label{jac_A}
J_A = \arraycolsep=1.4pt\def\arraystretch{2.0}
\left(
\begin{array}{c c}
	\cfrac{-1+JX}{\tau} \phantom{1234}& 0 \\
	K(1-X) \phantom{1234}& - \cfrac{1}{\tau_f}\\
\end{array}
\right).
\eeq
With our parameters (Table \ref{tableParam}) the eigenvalues
 $\lambda_1 = \cfrac{JX-1}{\tau} \approx -12.6$ and $\lambda_2 = -\cfrac{1}{\tau_f} \approx -1.11$ are both negative confirming $A$ is an attractor.
\paragraph{Saddle-point $S$} The second critical-point is $S_1 (h_1 \approx 8.07; x_1 \approx 0.28)$.
Its eigenvalues are $\lambda_1 \approx -5.73$ and $\lambda_2 \approx 1.43$. It is a saddle-point.
\paragraph{Attractor $A_2$} The third critical-point is $A_2  (h_2 \approx 28.8; x_2 \approx 0.53)$.
Its eigenvalues are $\lambda_1 \approx -11.9$ and $\lambda_2 \approx -1.33$. It is another attractor. The two attractors are separated by the 1D stable manifold of the saddle-point $S_1$ (fig. \ref{application}A, solid black curve).\\
The phase-space of system \eqref{2Dsyst}, restricted to the region $\{x \leq 0.5 \mbox{ and } h \leq 30\}$
has the same topological properties than system \eqref{Drift}: one attractor and one saddle-point, the separatrix delimiting the basin of attraction is the stable manifold of $S_1$ (fig. \ref{application}A). The escaping trajectories exits and re-enters the basin of attraction several times before eventually escaping (fig. \ref{application}A, orange).\\
Thus, we can now understand that the interburst intervals correspond to the exit times
of trajectories from the basin of attraction. Using formula \eqref{pdfExitTimes} to fit the distribution of exit times, we obtain that $\tilde{p} \approx 0.13$ (fig. \ref{application}B) and
\beq\label{f0bursts}
f_0(t) = 0.23\exp(-0.25t)\left(1+erf\left(\cfrac{t-2.45}{0.43}\right)\right)
\eeq
and
\beq\label{f1bursts}
f_1(t) = 0.19\exp(-0.25t)\left(1+erf\left(\cfrac{t+15.97}{0.58}\right)\right).
\eeq
Finally, similar to the generic system \eqref{Drift}, the RT number before escape does not
depend on the noise amplitude (fig. \ref{application}C). Trajectories are making on average 8 RT before escape (inset). To determine the mean escape time, we use formula \eqref{tEscSummed} and obtain   $\langle\tau_{esc}\rangle \approx \langle\tau_0\rangle + 7.7(\langle\tau_{ext}\rangle+\langle\tau_{int}\rangle)$ where $\langle\tau_0\rangle \approx 4.35s$ and $\langle\tau_{ext}\rangle+\langle\tau_{int}\rangle \approx 0.7s$ (fig. \ref{application}D) thus leading to a factor 2.2 in the escape time.
%%%%%%%%%%%%%%%%%%%%%%%%%%%%%%%%%%%%
%%%%%%%%%%%%%%%%%%%%%%%%%%%%%%%%%%%%
At this stage we conclude that long interburst durations, generated by excitatory neuronal networks \cite{Rouach_CxKO}, can be explained by the recurrent escape mechanism introduced here.
\\
%%%%%%%%%%%%%%%%%%%%%%%%%%%%%%%%%%%%%%%%%%%%%%%%%
{\bf Concluding remarks:}
%%%%%%%%%%%%%%%%%%%%%%%%%%%%%%%%%%%%%%%%%%%%%%%%%
We presented an escape mechanism for which reaching the boundary of the deterministic
basin of attraction induced by noise is not sufficient to escape. After crossing the separatrix, the noise tends to bring trajectories back inside the basin of attraction until they reach a region  (escape cone-like domain $C_\infty$), narrow near $S$ and that widens with the distance. The size of the characteristic distance from $S$ (boundary layer) after which trajectories escape is $~\sqrt{\cfrac{\lambda_+}{\sigma}}$ \cite{Schuss:Book}. We derived formulas for the mean escape time and the distribution of escape times taking into account the excursions inside and outside of the basin of attraction  before the final escape. \\
%%%%%%%%%%%%%%%%%%%%%%%%%%%%%%%%%
\begin{table}
	\begin{tabular}{l l l}
		& Parameters & Values \\
		\hline
		$\tau$ & Time constant for $h$ & 0.05s \cite{AHPmodel}\\
		$J$ & Synaptic connectivity & 4.21 \cite{AHPmodel}\\
		$K$ & Facilitation rate & 0.037Hz \cite{AHPmodel}\\
		$X$ & Facilitation resting value & 0.08825 \cite{AHPmodel}\\
		$L$ & Depression rate & 0.028Hz \cite{AHPmodel}\\
		$\tau_r$ & Depression time rate & 2.9s \cite{AHPmodel}\\
		$\tau_f$ & Facilitation time rate & 0.9s \cite{AHPmodel}\\
		$T$ & Depolarization parameter & 0 \\
		\hline
	\end{tabular}
 	\caption{Model \eqref{sys} parameters}\label{tableParam}
\end{table}
%%%%%%%%%%%%%%%%%%%%%%%%%%%%%%%%%%%%
%%%%%%%%%%%%%%%%%%%%%%%%%%%%%%%%%%%%%%%%%%%%%%%%%
% BIBLIOGRAPHY %%%%%%%%%%%%%%%%%%%%%%%%%%%%%%%%%%%%%%%%%%%%%%%%%%%%%%%%%%%%
%\bibliographystyle{IEEEtranlz}
\bibliography{biblio2}
\end{document}